\documentclass[preprint,12pt]{elsarticle}

\usepackage{amssymb}

\usepackage{graphicx,color}
\usepackage[usenames,dvipsnames]{xcolor}
\usepackage[section]{placeins}
\usepackage[normalem]{ulem}

\journal{Physics Letters B}

\begin{document}

\begin{frontmatter}
\title{Bethe-Salpeter scattering amplitude in Minkowski space}
\author{J.~Carbonell}
\address{Institut de Physique Nucl\'eaire,
Universit\'e Paris-Sud, IN2P3-CNRS, 91406 Orsay Cedex, France}
\author{V.A.~Karmanov}
\address{Lebedev Physical Institute, Leninsky Prospekt 53,
119991 Moscow, Russia}

\begin{abstract}
The off-mass shell scattering amplitude, satisfying the Bethe-Salpeter equation for spinless particles in Minkowski space with the ladder kernel, is computed for the first time.
\end{abstract}
\end{frontmatter}

\section{Introduction} \label{intro}

Obtaining the solutions of  Bethe-Salpeter (BS) equation in its original Minkowski space formulation \cite{bs} has raised an increasing interest in the recent years \cite{Kusaka:1997xd,bs1,bs2,Sauli_JPG80,fsv-2012}.
On one hand,  the Wick rotation itself is not directly applicable
for computing electromagnetic form factors \cite{ckm_ejpa,ck-trento} due to the existence of  singularities in the complex momentum plane whose contributions are in general unknown.
The Euclidean solutions are still used  in the context of BS-Schwinger-Dyson equations and  it is claimed they provide reliable results for bound state form factors at the price of a numerical {\it tour de force}
\cite{Oettel,Bhagwat:2002tx,Maris:2005tt,Krassnigg:2009gd,Eichmann,Strauss:2012dg,Windisch:2012sz}.
On the other hand, the off-shell BS scattering amplitude --
mandatory for important physical applications --  like computing  the transition e.m. form factor $\gamma^*d\to np$, or  solving the three-body BS-Faddeev equations --
requires a full Minkowski solution which  has not yet been obtained.

A method based on the  Nakanishi  representation  \cite{nak63} of the BS amplitude was developed in \cite{bs1,bs2}   allowing
to compute for the first time the bound state Minkowski amplitude and latter on \cite{ckm_ejpa,ck-trento} the corresponding  form factors. Although this approach  could be naturally extended to the scattering states,
we have found  that the problem  could be  solved  in a simplest and more straightforward way.
The aim of this paper is to presents  a direct solution of the original BS equation
in Minkowski space providing the scattering length,  elastic and inelastic phase shifts and the first results for the half-off-shell BS amplitude.

\section{Method for solving the equation}

\begin{figure}[h!]
\centering
\includegraphics[width=11.cm]{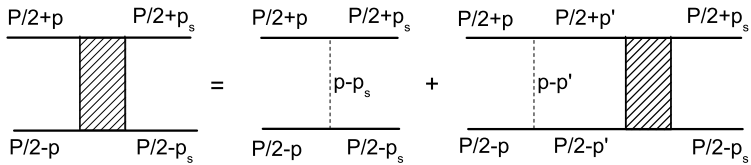}
\caption{Bethe-Salpeter equation for a scattering state.}\label{bs_eq}
\end{figure}

The inhomogeneous BS equation is graphically represented in figure \ref{bs_eq}.  In Minkowski space it reads:
\begin{equation}\label{BSE}
F(p,p_s;P)=K(p,p_s;P)- i\int\frac{d^4p'}{(2\pi)^4}
\frac{K(p,p';P) F(p',p_s;P)}
{\left[\left(\frac{P}{2}+p'\right)^2-m^2+i\epsilon\right]
\left[\left(\frac{P}{2}-p'\right)^2-m^2+i\epsilon\right]}
\end{equation}
We denote by $p$ the relative (off-shell) four-momentum variable of the equation, $p_s$  the
scattering momentum and $P$  the total momentum of the state with $P^2=M^2$,  the squared total mass of the system.
We will  consider hereafter  spinless particles ($m$=1) interacting by the one-boson exchange kernel $K$:
\begin{equation}\label{obe}
K(p,p';P)=-\frac{16\pi m^2\alpha}{(p-p')^2-\mu^2+i\epsilon},
\end{equation}
where $\alpha$  is the dimensionless coupling constant  of the Yukawa potential $V(r)=-\alpha\exp(-\mu r)/r$.

The difficulty in computing  the off-shell amplitude $F(p,p_s;P)$ in the entire domain of its arguments is due the singular character of the amplitude itself
as well as in the integrand of  the BS equation in Minkowski space (\ref{BSE}). These singularities are integrable in the mathematical sense, due to $i\epsilon$ in the denominators of propagators,
but  their  integration is a quite delicate task and requires the use of appropriate analytical as well as  numerical methods.

To avoid these problems, equation (\ref{BSE}) was first solved on-shell  \cite{tjon}  by rotating the integration contour $p_0\to ip_4$ and taking into account
the contributions of the crossed singularities,  which are absent in the bound state case.
Few other methods were also developed in the same line \cite{Schwartz_Morris_Haymaker}.
Until now, the off-shell amplitude has been computed only for a separable kernel \cite{burov}.
A method similar to the one developed in \cite{bs1,bs2} has been proposed in \cite{fsv-2012} for solving the scattering states  although  the numerical solutions are not yet available.

The amplitude $F(p,p_s;P)$ depends on the three four-momenta $p,p_s,P$. For a given incident  momentum $\vec{p}_s$ and written in the center of mass frame $\vec{P}=0$, $P_0=M=2\varepsilon_{p_s} =2\sqrt{m^2+p_s^2}$,
$F$  depends on three scalar variables  $|\vec{p}|$, $|\vec{p_s}|$ and $z=\cos(\vec{p},\vec{p}_s)$.
It will be hereafter denoted  by $F(p_0,p,z;p_s)$, setting abusively $p=|\vec{p}|$, $p_s=|\vec{p_s}|$.

In this letter we will restrict  to sketch our solution method and to present the first results for the S-wave on-mass  and half-off-mass shell amplitude
$F_0(p_0,p)$ obtained by the following partial wave decomposition of eq. (\ref{BSE}):
\begin{equation} 
 F_0(p_0,p)=\frac{1}{32\pi}\int_{-1}^1dz \; F(p_0,p,z;p_s)
 \end{equation}

The amplitude we consider here is a particular case of the so called full off-shell amplitude $F_0(p_0,p,p_{0s},p_s;M)$.  
The latter, in addition to the variables $p_0,p$ depends also on the off-shell independent variables $p_{0s},p_s$, 
now with $p_{0s}\neq \varepsilon_{p_s}$. The total mass $M=\sqrt{s}$ is neither equal to $2\varepsilon_{p_s}$ nor related to $p_{0s}$. 
By "off-shell amplitude" we will hereafter mean half-off-shell amplitude. 
The method we have developed  is also applicable to the full off-shell amplitude, 
though its dependence on two extra variables $p_{0s},p_s$ requires much more extensive numerical calculations.

The on-shell amplitude $F^{on}_0\equiv F_0(p_0=0,p=p_s)$ determines the phase shift according to:
\begin{equation}\label{delta}
\delta_0=\frac{1}{2i}\log\Bigl(1+\frac{2i p_s } {\varepsilon_{p_s}} F^{on}_0\Bigr)
\end{equation}
Several steps must be accomplished  before obtaining a
soluble equation for $F_0$ which takes into account  the four sources of singularities of the BS equation:

{\it (i)} The propagators in the  r.h.-side of  (\ref{BSE}) have two poles, each of them represented as sum of principal value and  $\delta$-function.
Their product gives rise to terms having respectively 0, 1 and 2 $\delta$'s. After partial wave decomposition, the 4D equation (\ref{BSE}) is reduced into a 2D one.
Integrating analytically  over $p'_0$ the $\delta$ contributions and eliminating the
principal values singularities by subtractions, one is left  with an S-wave equation in the form:
\begin{eqnarray}
&&F_0(p_0,p)  =  F^B_0(p_0,p)    +  \frac{i\pi^2 p_s}{8\varepsilon_{p_s}} W_0^S(p_0,p,0,p_s) F_0(0,p_s)
 \cr
                 &+& \frac{\pi}{2M} \int_0^{\infty}  \frac{dp'}{ \varepsilon_{p'} ( 2\varepsilon_{p'}-M) }    \left[{p'}^2 W_0^S(p_0,p, a_-,p') F_0(| a_- |,p')
     -\frac{2 {p_s}^2\varepsilon_{p'}}{\varepsilon_{p'} +\varepsilon_{p_s}}W_0^S(p_0,p,0,p_s) F_0(0,p_s)\right]
\cr
                    &-&    \frac{\pi}{2M} \int_0^{\infty} \frac{{p'}^2 dp'}{\varepsilon_{p'} (2\varepsilon_{p'}+M) }      W_0^S(p_0,p, a_+,p') F_0(a_+,p' )   \cr
                 &+&  \frac{i}{2M}  \int_0^{\infty}  \frac{{p'}^2dp' }{\varepsilon_{p'}}   \int_0^{\infty} dp'_0 \left[ \frac{ W^S_{0}(p_0,p,p'_0,p') F_0(p'_0,p')  - W^S_{0}(p_0,p,a_-,p')  F_0(|a_- |,p')}{  {p'}_0^2-a_-^2 }\right]  \cr
                &-&    \frac{i}{2M}  \int_0^{\infty} \frac{{p'}^2dp'} {\varepsilon_{p'}}     \int_0^{\infty} dp'_0 \left[ \frac{ W^S_{0}(p_0,p,p'_0,p') F_0(p'_0,p')  - W^S_{0}(p_0,p,a_+,p') F_0(a_+,p')}  {{p'}_0^2-a_+^2 } \right]
                   \label{Eq_F_sym}
\end{eqnarray}
where $a_{\mp} =\varepsilon_{p'} \mp \varepsilon_{p_s}$ and $W_0^S$ is the S-wave kernel -- suitably symmetrized on $p'_0$ variable to restrict its integration domain to $[0,\infty]$ --
is given by
\[ W_0^S(p_0,p,p'_0,p')=W_0(p_0,p,p'_0,p')+W_0(p_0,p,-p'_0,p')\]
with:
\begin{equation}\label{W0}
W_0(p_0,p,p'_0,p') = -\frac{\alpha m^2}{\pi pp'} \left\{  \frac{1}{\pi}  \log \left| \frac{(\eta+1)}{(\eta-1)} \right| - i I(\eta)   \right\},
\quad
I(\eta)=\left\{
\begin{array}{lcrcl}
1  & {\rm if} & \mid\eta\mid &\leq& 1 \cr
0  & {\rm if} & \mid\eta\mid &> & 1
\end{array}\right.
\end{equation}
and
\[  \eta =   {1\over2pp'} \left[ (p_0  - p'_0)^2 - p^2 - {p'}^2 -\mu^2  \right]  \]

The inhomogeneous (Born)  term $F^B_0$ reads:
\[ F^B_0(p_0,p)={\pi^2\over 4}W_0(p_0,p,0,p_s)\]

The details of the derivation of eq. (\ref{Eq_F_sym}) as well as its generalization to the full off-shell amplitude  are quite lengthy  and will be given in a forthcoming publication. 
The origin of the different terms appearing in (\ref{Eq_F_sym})
are however quite clear. The non-integral term in the first line, 
follows from the integrated (2D) product of the two $\delta$-function   mentioned above.
The one-dimensional integrals -- second and third lines  --  results from  one $\delta$-function terms, after integration over $p'_0$. 
The last two lines come from the principal values (PV) alone. 
The differences appearing in the squared brackets correspond to removing the pole singularities at $2\epsilon_{p'}=M$   (second line) and $p'_0=a_{\pm}$
(third and forth lines) according to the well known subtraction technique eliminating singularity:
$$
PV\int_0^{\infty} \frac{f(x')dx'}{{x'}^2-a^2}=\int_0^{\infty} dx' \left[\frac{f(x')-f(a)}{{x'}^2-a^2}\right]
$$
In l.h.-side the integrand  at $x'=a$ is singular that complicates the numerical calculation of integral, whereas r.h.-side does not contain this singularity.

{\it (ii)} The  propagator of the exchanged particle (\ref{obe}) has  two poles which, after  partial wave decomposition, turn into logarithmic singularities in kernel (\ref{W0}).
Their positions are found analytically and the numerical integration over $p'_0$  is split into intervals between two consecutive singularities. Inside each of these intervals
an appropriate change of variable is made to make regular the integrand of eq. (\ref{Eq_F_sym}). We proceed in a similar way for the $p'$ integration.

{\it (iii)}  The inhomogeneous (Born) term $F_0^B$  has also logarithmic singularities  in both variables $p_0$, $p$
which  are analytically known.

{\it (iv)}  The amplitude $F_0$ itself has many singularities, among which those originated by the Born term $F_0^B$ are the strongest ones.
This makes difficult  representing $F_0$ on a basis of regular functions.
To circumvent this problem we made the replacement
$F_0=F_0^B f_0$, where $f_0$ is a smoother function. After that, the singularities of the inhomogeneous Born term are casted into the kernel
and integrated using the same procedure than in {\it (ii)}.

We obtain in this way a non-singular equation for  $f_0$ which we solve by standard methods. The  off-mass shell BS amplitude $F_0$ in Minkowski space is thus safely computed.

\section{Numerical results}
\bigskip\bigskip

\begin{figure}[h]
\centering
\includegraphics[width=11.cm]{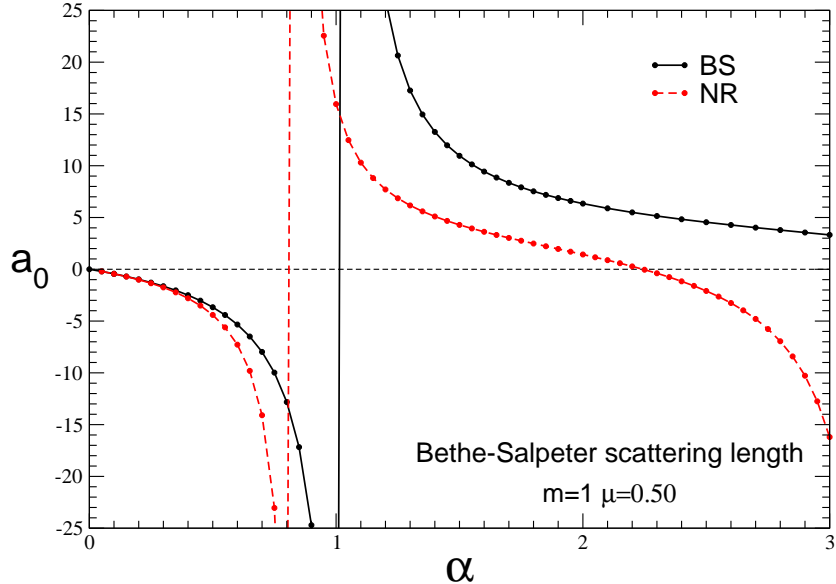}
\caption{BS scattering length $a_0$ versus the coupling constant  $\alpha$ (solid), compared to the non-relativistic results (dashed) for $\mu=0.5$.}\label{fig_a0}
\end{figure}

Our first check was to solve the bound state problem by dropping the inhomogeneous term in (\ref{BSE}) and setting $M=2m-B$.
The binding energy $B$  thus obtained,  coincides within four-digit accuracy, with the one calculated, by other method, in our previous work \cite{bs1}.

The S-wave off-shell scattering amplitude $F_0$ was then  calculated and the phase shifts extracted according to (\ref{delta}).
An independent equation, similar to one obtained in \cite{tjon}, have also been  derived,
which  provides the phase shifts and the Minkowski off-shell amplitude  for the particular value $p_0=\varepsilon_{p_s}- \varepsilon_{p}$ restricted to the interval $0\le p \le p_s$.
The values  found by these two independent methods are consistent to each other.

The BS relativistic formalism accounts naturally for  the meson creation in the scattering process, when the available  kinetic energy allows it.
The inelasticity threshold corresponding to the $n$-particle creation is given by
\begin{equation}
Êp^{(n)}_s = m \sqrt{  \left( {\mu\over m}\right)  \;n  +    {1\over4}\left( {\mu\over m}\right) ^2 \; n^2 }
\end{equation}
Below the first inelastic threshold,  $p^{(1)}_s=\sqrt{m\mu+\mu^2/4}$, the phase shifts are real.
This unitarity condition is not automatically fulfilled in our approach, but appears as a consequence of handling the correct solution
and provides a stringent test of the numerical method.
Above  $p^{(1)}_s$, the phase shift obtains an imaginary part which behaves like
\begin{equation}
 {\rm Im}(\delta_0)\sim (p_s-p^{(1)}_s)^2
 \end{equation}
in the threshold vicinity.
Higher inelasticity thresholds, corresponding to creation of 2, 3, etc. intermediate mesons at $p^{(n)}_s$, are also  taken into account in our calculations.

\begin{table}[h]
\begin{center}
\caption{Scattering length values obtained with Bethe-Salpeter equation  as a function of the coupling
constant $\alpha$ for $m=1$ and different values of the exchanged mass   $\mu=0.15$ , $\mu=0.50$ and $\mu=1.00$.}\label{tab_a_alpha}
\bigskip

\begin{tabular}{l r  r  r}\hline
$\alpha$      &    $\mu=0.15$  &   $\mu=0.50$     &  $\mu=1.00$  \\\hline
0.01              & -0.460D+00     &  -0.403D-01       &  -0.100D-01              \cr
0.05              & -0.270D+01     &  -0.209D+00      &  -0.510D-01               \cr
0.10              & -0.692D+01     &  -0.438D+00      &  -0.104D+00 \cr
0.20              & -0.346D+02     &   -0.971D+00     &   -0.217D+00		 \cr
0.30              &  0.795D+02     &  -0.164D+01       &   -0.339D+00   \cr
0.40              &  0.272D+02     &   -0.250D+01      &   -0.474D+00	 \cr
0.50              &  0.214D+02     &    -0.366D+01     &   -0.621D+00\cr
0.60              &  0.128D+02     &    -0.534D+01     &    -0.784D+00 	 \cr
0.70              &  0.866D+01     &    -0.798D+01     &   -0.965D+00  \cr
0.80              &  0.373D+01     &     -0.128D+02    &     -0.117D+01	\cr
0.90              & -0.457D+01    &   -0.247D+02       &     -0.140D+01  \cr
1.00              & -0.281D+02    &   -0.103D+03       &   -0.166D+01   \cr
1.10              &  0.900D+03    &  0.620D+02         &   -0.195D+01  \cr
1.50              &  0.247D+02    &    0.110D+02        &   -0.379D+01		 \cr
2.0                &  0.174D+02     &    0.634D+01        &   -0.111D+02  	  \\
2.5                &   0.144D+02    &    0.454D+01       &   0.568D+02  	  \\\hline
\end{tabular}
\end{center}
\end{table}

The low energy parameters were computed
and found to be consistent with a quadratic fit to the effective range function $p\cot\delta(p)= -\frac{1}{a_0} + \frac{1}{2} r_0 p^2 $ .
The BS scattering length  $a_0$ as a function of the coupling constant $\alpha$ is given in figure \ref{fig_a0} for  $\mu=0.50$.
It is compared to the non-relativistic (NR)  values provided by the Schr\"odinger equation with the Yukawa potential.
The singularities correspond to appearance of the first bound state at $\alpha_0=1.02$ for BS and  $\alpha_0=0.840$ for NR.
As one can see, the differences between a relativistic and a non-relativistic treatments of the same problem
are not of kinematical origin since even for  processes involving zero energy they  can be substantially large, especially in presence of bound state.
It is worth noticing that only  in  the limit $\alpha\to0$  the two curves are tangent to each other
and in this region the results are given by the  Born approximation
\begin{equation}
a^B_0=   - \; {1\over\mu} {m\over\mu} \alpha
\end{equation}
which is the same for the NR and the BS equation.
Beyond this region both dynamics are not compatible.
Some selected  numerical values of the scattering length are listed in Table \ref{tab_a_alpha} for different values of  the exchanged mass $\mu$.
As one can see by direct inspection, the scaling properties of the non relativistic equation \cite{SCQ_EPJA47_2011}, in particular  the relation between the
scattering length corresponding to different values of $\mu$ and coupling constants
\begin{equation}
 a_0\left( {\mu\over m} ,\alpha\right) = {1\over \mu} a_0\left(1,{\alpha\over {\mu\over m}} \right)
\end{equation}
are no longer valid except in the Born approximation region.

\vspace{0.5cm}
\begin{figure}[ht!]
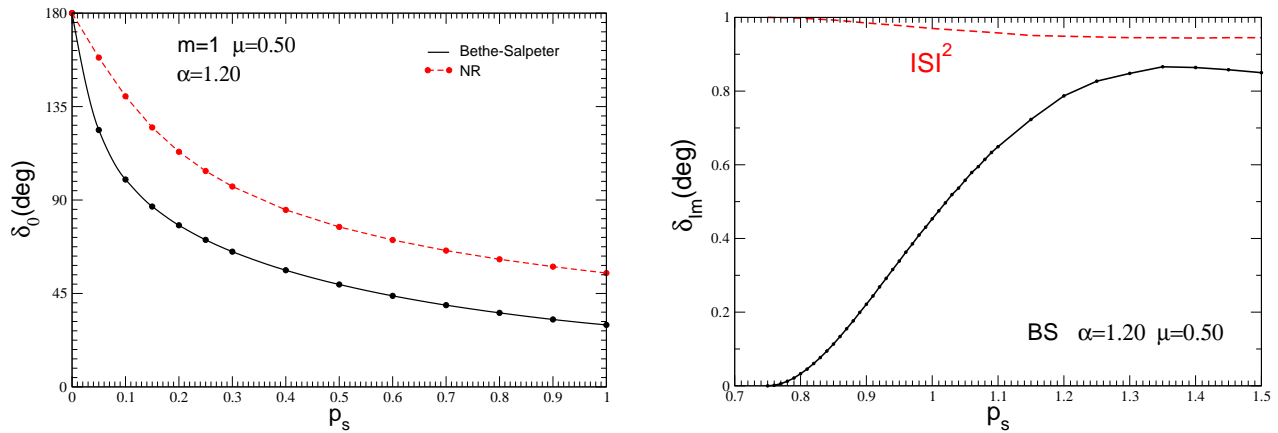

\begin{center}
\includegraphics[width=8cm]{Phaseshifts_alpha_1.20_0.50.eps}
\hspace{0.5cm}
\includegraphics[width=8cm]{Inel_alpha_1.20_mu_0.50.eps}
\end{center}
\caption{Real (left panel) and imaginary (right panel)  phase shift (degrees) for $\alpha=1.2$ and $\mu=0.50$ calculated  via BS equation (solid)  compared to the non-relativistic results (dashed).}\label{fig2}
\end{figure}

\begin{table}[h]
\caption{Real and imaginary parts of the phase shift (degrees) calculated by  BS eq. (\ref{BSE}) vs. incident momentum $p_s$ for $\alpha=1.2$  and $\mu=0.5$.
Corresponding first inelastic threshold is $p^{(1)}_s=0.75$}
\label{tab2}
\begin{center}
\begin{tabular}{lllll ll   l l l  l  l    ll}
\hline
$p_s$            & 0.05  & 0.1      & 0.2   &  0.3     & 0.4     &   0.5   &   0.6     &  0.7    & 0.8      &  0.9         & 1.0       & 1.3      & 1.5\\ \hline
$Re[\delta]$  & 124  &  99.9   & 77.8 &   65.1  & 56.2  &  49.3  &  43.9   &  39.4  & 35.7    &   32.5     & 29.7    &  22.8   &  19.3\\
$Im[\delta]$  & 0        &      0     &  0      &     0     & 0       &      0    &     0      &      0    & 0.033  &   0.221             & 0.453  & 0.848   & 0.852 \\ \hline
\end{tabular}
\end{center}
\end{table}

Figure \ref{fig2}  (left panel) shows the real  phase  shifts calculated with BS  (solid line)  and NR (dashed line) equations and the same parameters than in fig. \ref{fig_a0}.
For this value of $\alpha$ there exists a bound state and, according to the Levinson theorem, the phase shift starts at 180$^\circ$.
One can see that the difference between relativistic and non-relativistic results is considerable even for relatively small incident momentum.
The right panel  shows the imaginary part of the phase shift. It appears  starting from the first inelastic meson-production threshold $p^{(1)}_s=0.75$ and displays the expected quadratic behavior.
Simultaneously the modulus squared of the S-matrix (displayed in dashed line) starts differing from unity.
The results of this figure contain  the contributions of the second $p^{(2)}_s=1.118$ and third  $p^{(3)}_s=1.435$ meson creation thresholds as well.
Corresponding numerical values are given in  table \ref{tab2}.

\begin{figure}[ht!]
\begin{center}
\includegraphics[width=8cm]{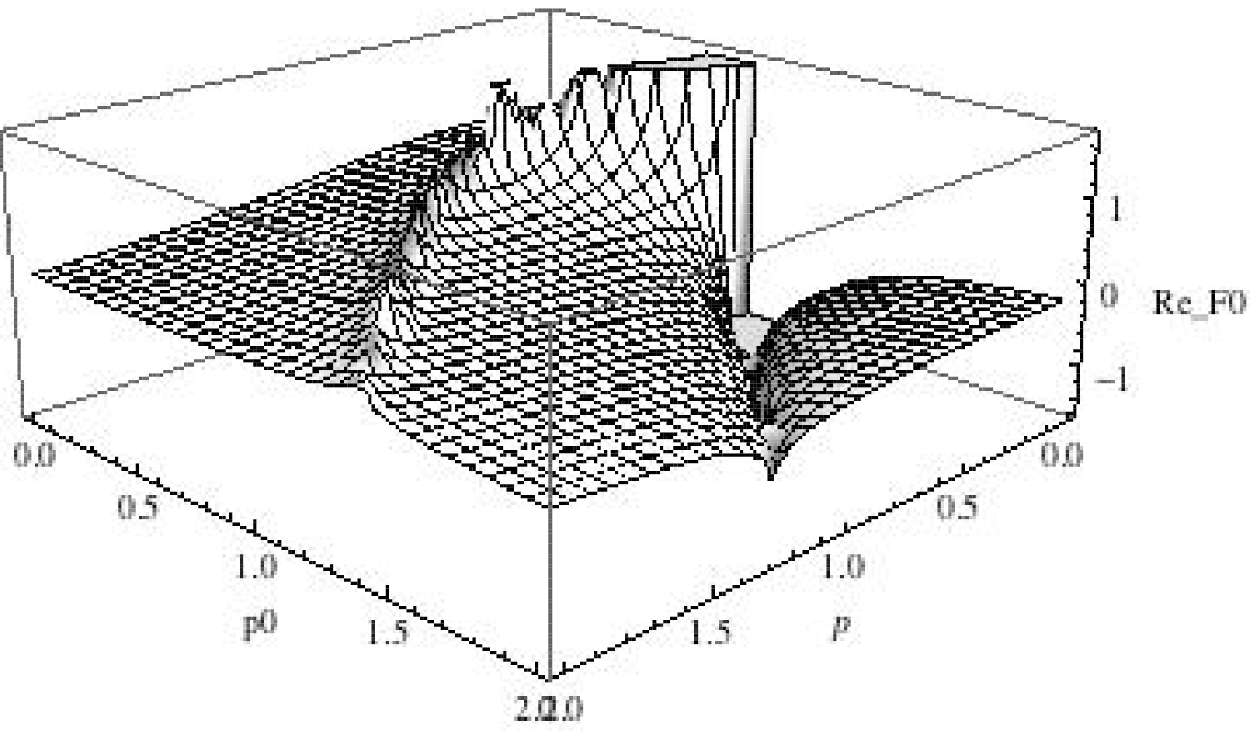}
\includegraphics[width=8cm]{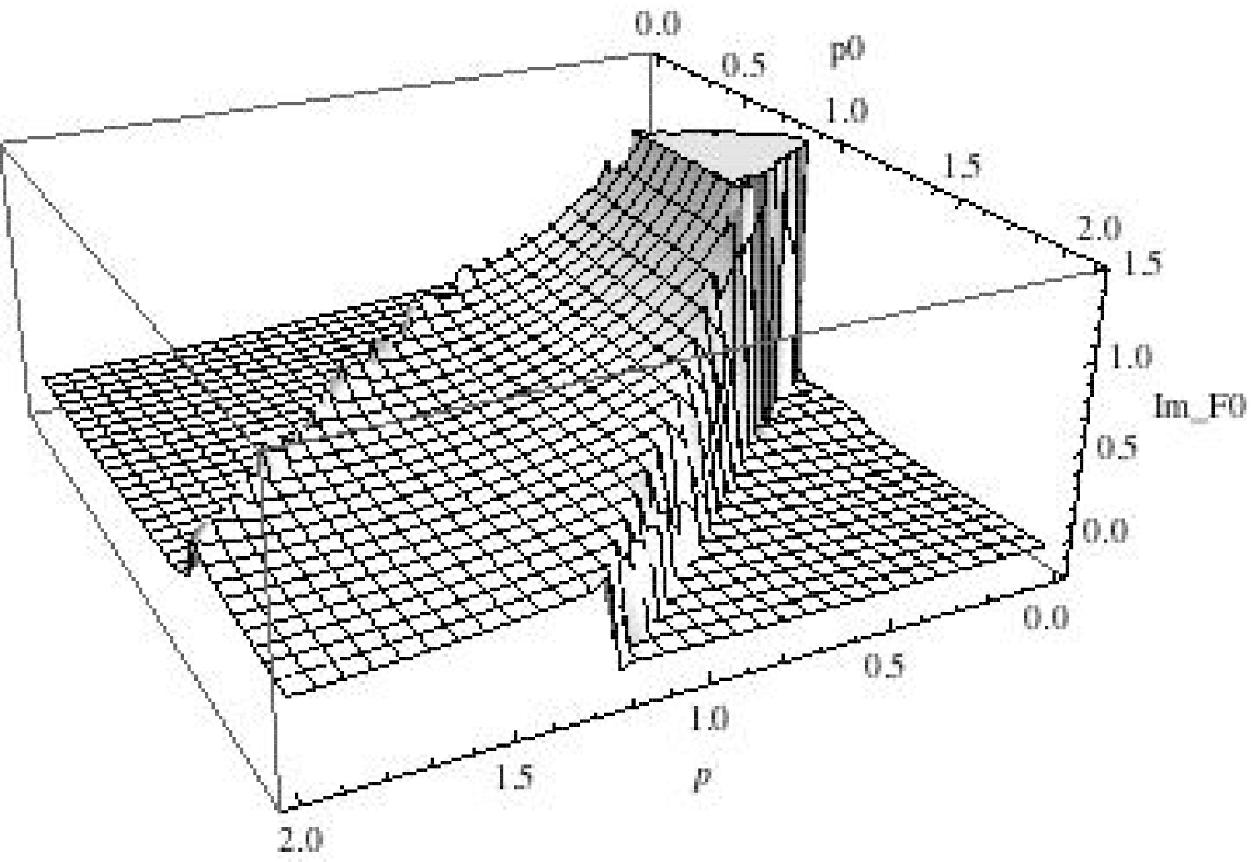}
\end{center}
\caption{Real  (left panel) and imaginary (right panel) parts of the off-shell amplitude $F(p_0,p;p_s)$ for $\alpha=0.5$, $p_s=\mu=0.5$.} \label{fig3}
\end{figure}
Finally, we display in Fig. \ref{fig3} the real (left panel) and imaginary (right panel) parts of the off-shell scattering amplitude $F_0(p_0,p)$
as a function of  $p_0$ and $p$ calculated for $\alpha=0.5$, $p_s=\mu=0.5$.
Its real part shows a non trivial structure with  a ridge and a gap resulting from the singularities of the inhomogeneous term.
Its on-shell value $F_0^{on}=F_0(0,p_s)=0.753+i0.292$,
determining  the phase shift $\delta=21.2^{\circ}$, corresponds to a single point on theses two surfaces.
Our calculation, shown in Fig. \ref{fig3}, provides   the full amplitude $F_0(p_0,p)$ in  a two-dimensional domain.

Computing this quantity, and  related on-shell observables, is the main result of this work.
Together with the bound state solution in Minkowski space \cite{bs1},  they pave the way for a consistent relativistic description of composite systems in the framework of  BS equation.

\section{Conclusion}
We have presented the first results of the BS off-shell scattering amplitude in Minkowski space.
The different kinds of singularities of the original BS equation are properly treated. A regular equation
is obtained and solved by standard methods.
The results presented here were limited  to S-wave in the spinles case and the ladder kernel
but they can be extended to any partial wave.
Coming on mass shell,  the elastic phase shifts where accurately computed.
They considerably differ, even at zero energy, from the non-relativistic ones.
Above the meson creation threshold, an imaginary part of the phase shift  appears  and has also been  calculated.
The  off-shell BS scattering amplitude thus obtained can be further used to calculate the transition form factor. 
In its full off-shell form, it can be used as  input in the three-body BS-Faddeev equations.


\end{document}